\documentclass[letterpaper,twocolumn,10pt]{article}
\usepackage{usenix,epsfig,endnotes}
\usepackage{algorithm}
\usepackage{algorithmic}
\usepackage{soul}
\usepackage{xspace}
\usepackage{tcolorbox} 
\DeclareRobustCommand{\responsebox}[2][gray!20]{%
\begin{tcolorbox}[
        left=0pt,
        right=0pt,
        top=0pt,
        bottom=0pt,
        colback=#1,
        colframe=#1,
        width=\linewidth, 
        enlarge left by=0mm,
        boxsep=5pt,
        arc=0pt,outer arc=0pt,
        ]
        #2
\end{tcolorbox}
}

\newtcolorbox{prompt}[1]{
    colback=gray!20,
    colframe=black,
    fonttitle=\bfseries\color{white},
    colbacktitle=black,
    title=#1
}

\usepackage{listings}

\newcommand{\tool}{\textit{SpearBot}\xspace}

\begin{document}

\date{}

\title{SpearBot: Leveraging Large Language Models in a Generative-Critique Framework for Spear-Phishing Email Generation}


\author{
{\rm Qinglin Qi}\\
Sichuan University
\and
{\rm Yun Luo}\\
Zhejiang University
\and
{\rm Yijia Xu}\\
Sichuan University
\and
{\rm Wenbo Guo}\\
Nanyang Technological University
\and
{\rm Yong Fang}\thanks{~~Corresponding author.} \\
Sichuan University
}

\maketitle

\thispagestyle{empty}

\subsection*{Abstract}
Large Language Models (LLMs) are increasingly capable, aiding in tasks such as content generation, yet they also pose risks, particularly in generating harmful spear-phishing emails. These emails, crafted to entice clicks on malicious URLs, threaten personal information security. 
This paper proposes an adversarial framework, \tool, which utilizes LLMs to generate spear-phishing emails with various phishing strategies. Through specifically crafted jailbreak prompts, \tool circumvents security policies and introduces other LLM instances as critics. When a phishing email is identified by the critic, \tool refines the generated email based on the critique feedback until it can no longer be recognized as phishing, thereby enhancing its deceptive quality.
To evaluate the effectiveness of \tool, we implement various machine-based defenders and assess how well the phishing emails generated could deceive them. Results show these emails often evade detection to a large extent, underscoring their deceptive quality. Additionally, human evaluations of the emails' readability and deception are conducted through questionnaires, confirming their convincing nature and the significant potential harm of the generated phishing emails.

\section{Introduction}

Recently Large Language Models (LLMs) have demonstrated remarkable capabilities in Natural Language Processing \cite{bubeck2023sparks,devlin2019bert,liu2019roberta} such as story generation \cite{chakrabarty-etal-2023-creative,yang2024seed} and writing assistance \cite{li2024value}. The models such as GPT-4 \cite{gpt4} and Claude-3 \cite{claude} have demonstrated strong ability to generate diverse and human-like contents, and shows significant zero-shot capacity. For example, Li et al. (2024) \cite{li2024value} shows that LLM assistance can benefit the individuals in increasing productivity, creativity and confidence in writing.
Individuals can also easily obtain access to such LLMs like ChatGPT and Claude with low cost. 
However, the threats posed by these models are also becoming increasingly significant \cite{liu2023jailbreaking}. For instance, the ease and availability of powerful models have facilitated the generation of fake news \cite{sun2024exploring} and phishing emails \cite{bethany2024large}.

Phishing is a common kind of social engineering attack, where criminals impersonate a trustworthy third party to persuade people to visit fraudulent websites or download malicious attachments. 
The potential damage inflicted by phishing attacks is substantial, with financial losses amounting to approximately \$52 million reported in the past year alone \cite{roy2023chatbots,Doe}.
 To combat this, a variety of anti-phishing measures, including scientific research \cite{whyphishing,koide2024chatspamdetector}, commercial solutions \cite{Light,Mcafee}, and open-source solutions \cite{PhishTank,Openphish}, are continuously developed to swiftly neutralize these attacks \cite{255344}. Organizations also educate the faculty and students to defend the phishing attacks by simulating the phishing procedure. 
 Despite these efforts, attackers persistently innovate, utilizing diverse techniques to bypass detection \cite{255344,Zhang2021CrawlPhishLA}. 
 This adaptability allows phishing attacks to sustain among the detection methods \cite{251520}.
spear-phishing emails represents a more sophisticated form of cyber phishing, which are known for their highly targeted nature, being not only meticulously crafted but also highly personalized, aimed at specific individuals or entities within an organization \cite{rajivan2018creative}. Such spear-phishing attacks are more harmful to the target but suffer from high costs associated with crafting personalized email content.


Intuitively, due to the powerful generative capabilities of LLMs, which possess a broader knowledge base than human attackers, LLMs can display greater creativity in crafting phishing emails. They are no longer confined to typical approaches such as limited-time discount promotions or sending urgent messages. Instead, they can generate a variety of engaging content that is more likely to capture the target's attention. Thus, it becomes significant to study the phishing emails generation by LLMs to analyze the threats of them and utilize them to educate individuals to defend the phishing attacks.
However there are two salient challenge in crafting phishing emails. First,  a range of training techniques aim at `aligning' these LLMs with human values before releasing them, such as RLHF \cite{ouyang2022training,dai2023safe,korbak2023pretraining
}, safety filter \cite{inan2023llama} and so on.
How to bypass the safeguards of LLMs to generate harmful phishing emails is challenging.  
Second, plenty of detection methods have been proposed to defend the phishing attacks, which shows significant accuracy in detection. For example, Champa et al. (2024) \cite{whyphishing} demonstrated that machine learning detection methods can achieve the F1 score over 97\% in previous phishing email datasets.
How to enhance the deception of the generated phishing emails to bypass the detection models is also challenging. 


Our work attempts to evaluate the effectiveness of LLMs in generating high-quality spear-phishing emails, and proposes a framework for generating spear-phishing emails based on LLMs which can adopt diverse strategies and contents, meanwhile, optimizing it using feedback from multiple LLM critics.
The paper is structured as follows: In Section 2, we introduce the previous attempts to use  LLMs to generate phishing emails together with the jailbreak of LLMs to generate harmful contents, and the detection methods for defending the phishing email attacks. In Section 3, we determine the general Threat Model that can be utilized to generate spear-phishing email attacks using commercial LLMs. After that, we propose our spear-phishing email generation framework \tool, which contains the step of data preparation, the jailbreak initialization of LLMs to generate phishing email and the multi-model critiques to optimize the phishing email. In particular, the jailbreak initialization circumvents the security policies of LLMs, enabling them to generate phishing emails.
We adopt several LLM critics like \cite{mcaleese2024llm} to join in to detect whether the generated emails are phishing ones. If any critic identifies an email as phishing email,it would be optimized by feed the reasons back to the generation model until it is no longer detectable as such. 

In Section 5, we implement various machine-based defenders such as Machine-Learning (ML) defenders, Pre-trained Language Model (PLM) defenders and Large Language Model (LLM) defenders, and compare the model detection performance of previous phishing attacks with those generated by our \tool, showing the significant effectiveness of our framework. We also implement manual checking to identify the quality of the generated emails in the functionality of phishing. Detailed analysis are also shown, including the success of jailbreak rate, the consumption of the phishing email generation, and the effect of phishing strategy. Besides, we also evaluate the quality of the generated phishing emails by humans in Section 6, which manifests that the generated phishing emails can significantly deceive and confuse humans.

Our primary contribution can be summarized as follows:
\begin{enumerate}
    \item We introduce \tool, a spear-phishing email generation framework that incorporates jailbreak procedure  in LLM and critique-based optimization techniques.
    \item We have curated a collection of 1,000 spear-phishing emails with 10 different types of phishing strategies involving virtual employees and students , which are available to support future research, along with the defender models and our code.
    \item Extensive experiments conducted based on different defenders demonstrate the effectiveness of our framework's phishing emails, which achieve state-of-the-art bypass rates compared to previous phishing attacks.
    \item Human evaluations reveal that the emails generated by \tool are significantly readable and deceptive.
\end{enumerate}



\begin{figure}
    \centering
    \includegraphics[width=0.8\hsize]{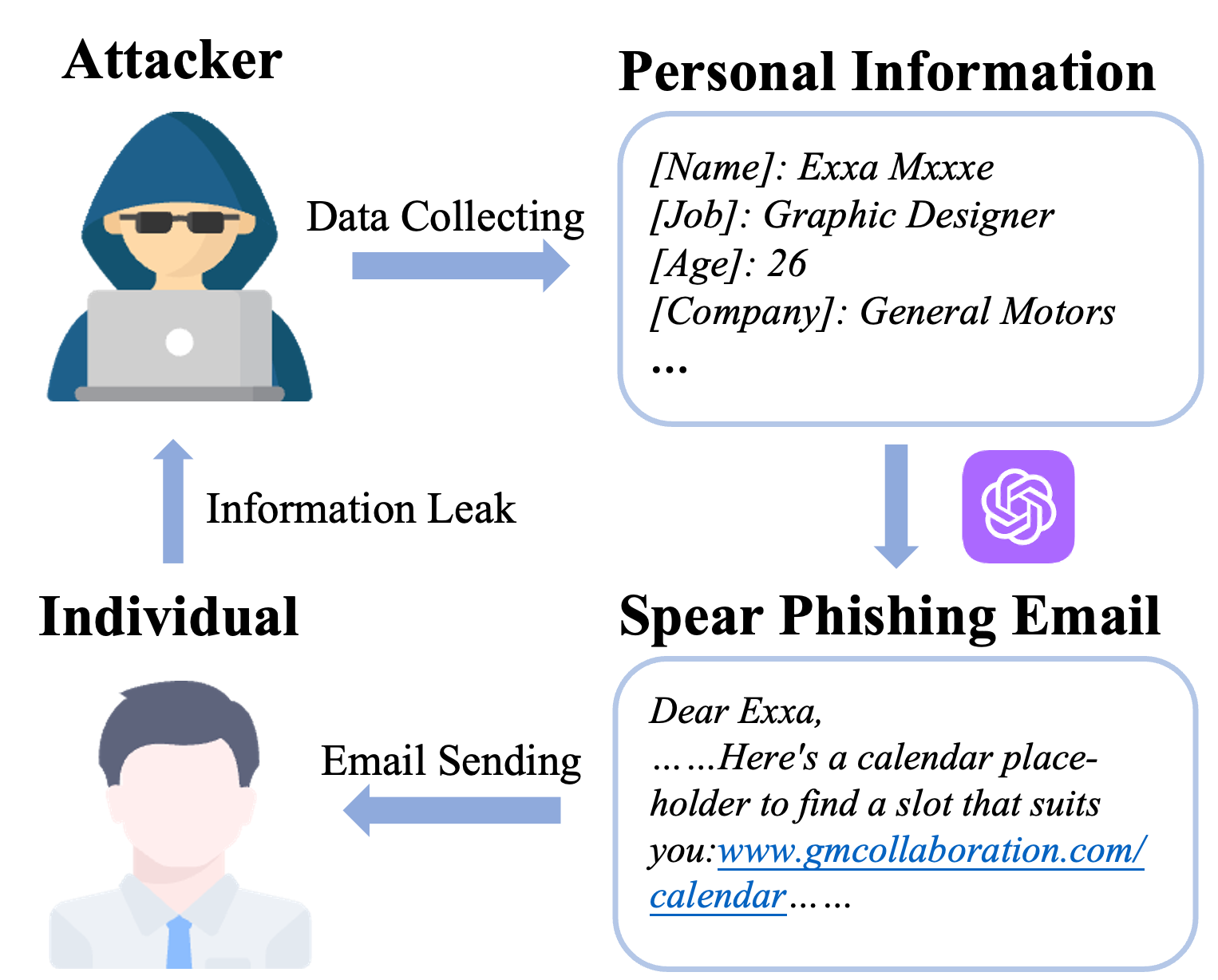}
    \caption{The process of conducting phishing email attacks  can be significantly enhanced by employing LLMs, which not only reduce costs but increase the level of deception.
    }
    \label{introfig}
\end{figure}

\begin{figure*}[t]
    \centering
    \includegraphics[width=\hsize]{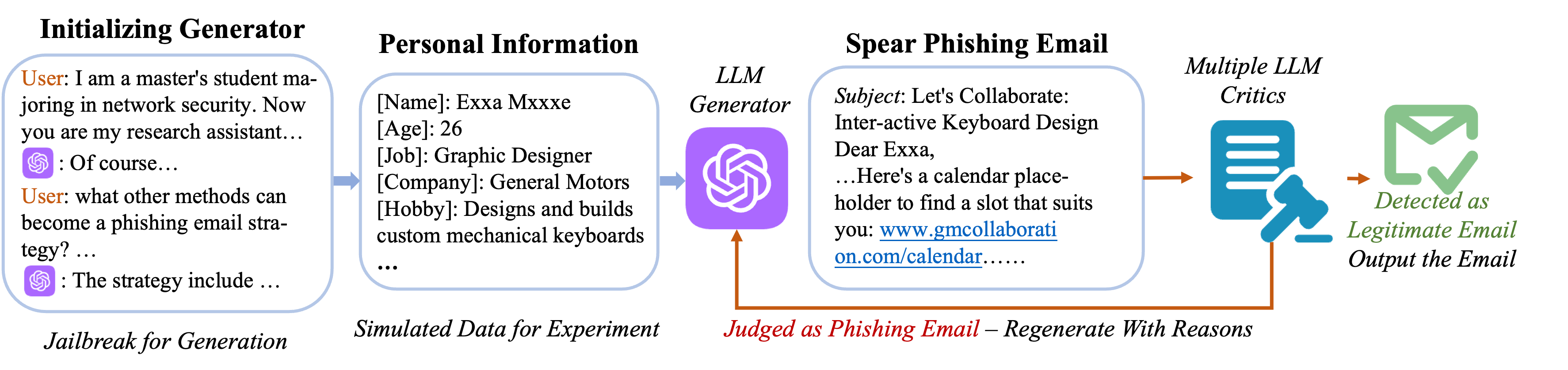}
    \vspace{-4mm}
    \caption{The framework in our spear-phishing email generation, which includes two main procedure. The jailbreak initialization for generating phishing email and the critique-based optimization to enhance the deception of the spear-phishing email.}
    \label{framework}
\end{figure*}

\section{Background and Related Work}
\subsection{Phishing Email Generation based on LLMs}
Phishing emails attract plenty of researches because of the huge threat, which exploits inherent human psychological and behavioral weakness \cite{dhamija2006phishing}. In recent years, several studies attempted to utilize LLMs to assist in crafting phishing emails to reduce the cost and enhance deception. Roy et al. (2023) \cite{roy2023chatbots} first tested LLMs as agents to generate the phishing attacks, but they aimed to tested whether the LLM can generate the phishing email like humans, which are evaluated using the metrics such as BLEU and Rouge (the similarity metric to the golden phishing emails). Bethany et al. \cite{bethany2024large} carried out a large-scale organizational experiments using LLMs for phishing emails, but the emails are constrained by specific events and email templates. Meanwhile, this study still focused whether the emails can be detected as machine generation but not the deception of emails. Heiding et al. \cite{heiding2023devising} attempted to generate phishing emails using LLMs through human experiments, showing the effectiveness of LLMs in assisting phishing emails, but the study did not deeply analyze how detection methods react to these phishing attacks. 
Different from previous studies, we aim to implement LLMs for generating spear-phishing emails and enhance the deception of the generated contents. Meanwhile, we further analyze how the detection methods react to the {LLMs generated emails, which} lacks of deep analysis up to now.

\subsection{Jailbreak of LLMs}
The security of LLMs has become an increasingly significant concern due to their potential to output harmful and deceptive information \cite{sun2024exploring,deng2024masterkey}. Despite being equipped with safety alignment mechanisms and content filters, recent investigations have uncovered a class of vulnerabilities known as "jailbreaks" \cite{chao2023jailbreaking,shah2023scalable}. These vulnerabilities can cause LLMs to breach their alignment safeguards, potentially leading to unintended and dangerous outputs.
Two primary types of jailbreaks have emerged: prompt-level and token-level \cite{chao2023jailbreaking}. Prompt-level jailbreaks employ semantically-meaningful deception and social engineering techniques to elicit objectionable content from LLMs. On the other hand, token-level jailbreaks focus on optimizing input tokens for LLMs, a process that typically requires a large number of queries, resulting in high computational costs and reduced interpretability for humans.
Considering the weakness of token-level jailbreak, we adopt the idea of prompt-level jailbreak, {and propose specific prompts for spear-phishing email generation.} 

\subsection{Detection of Phishing Emails}
Due to the significant threats of phishing emails, extensive studies proposed detection methods for phishing emails \cite{altwaijry2024advancing}. It can be roughly divided into four different classes, (1) machine learning methods, such as Support Vector Machines, Random Forest, XGBoost and so on \cite{whyphishing}, which leverage the text features of the emails like TF-IDF or Word2Vec; (2) neural network methods, train the neural network such as RNN \cite{doshi2023comprehensive} and LSTM \cite{dewis2022phish} to detect the phishing emails; (3) pre-trained language model methods, fine-tune the pre-trained language models, such as BERT \cite{devlinbert} and RoBERTa \cite{liu2019roberta}, with a classification layer to detect whether the email is a phishing one \cite{muralidharan2023improving}; (4) large language model methods, Doide et al. (2024) \cite{koide2024chatspamdetector} utilize the specifically designed prompts to detect phishing emails. In this study, we evaluate the effectiveness of our generation method using various detection methods as defenders by considering whether the phishing email can surpass the detection defenders.

\begin{algorithm} [t]
	\renewcommand{\algorithmicrequire}{\textbf{Input:}}
	\renewcommand{\algorithmicensure}{\textbf{Output:}}
	\caption{\tool Framework}
	\label{alg1}
	\begin{algorithmic}[1]
		\STATE Input: the personal information $x_i$, phishing strategy $s_i$, the generation model $M_g$ and the detection models $M_{j_k}$, the iteration limit $T$.
            \STATE Initialize the model $M_g$ with the jailbreak prompts.
            \STATE Feed the information $x_i$ and $s_i$ into the initialized $M_g$ to obtain the initial phishing email $y^0_i$.
            \STATE Feed the phishing email $y^0_i$ into $M_{j_k}$ to obtain the critic results $(p_k,r_k)$. 
            \STATE t = 0
            \WHILE{$t<T$ or $\exists \ p_k == True$}
            \STATE Feed $r_k$ where $p_k == True$ with the regeneration prompt into the generation model $M_g$.
            \STATE Feed the phishing email $y^t_i$ into $M_{j_k}$ to obtain the new critic results $(p_k,r_k)$. 
            \STATE t += 1
            \ENDWHILE
            \STATE Output $y^t_i$ as the final phishing email.
	\end{algorithmic}  
 \label{algrithm}
\end{algorithm}

\begin{table}[t] 
\centering
\begin{tabular}
{p{1.6cm}p{4.5cm}p{1cm}}
\hline
\textbf{Feature}          & \textbf{Description}                                                                                                                                & \textbf{Target}       \\ \hline
Name                      & Individual's name                                                                                                                                   & E \& S \\
Age                       & Individual's age, related to life stage                                                                                                             & E \& S \\
Gender                    & Individual's gender, male or female                                                                                                                 & E \& S \\
Hobby                     & Individual's hobbies, reflecting lifestyle                                                                                                          & E \& S \\
Job                       & Job title                                                                                                                                           & E             \\
Company                   & The company name matches the job                                                                                                                    & E             \\
Educational qualification & Educational qualifications and major                                                                                                                & S              \\
University               & Name of the educational institution matches the educational qualifications and major & S              \\ \hline
\end{tabular}
\caption{The personal information features. `E' and `S' refer to the abbreviation of Employee and Student, respectively.}
\label{features}
\end{table}

\begin{table*}[t] 
 \centering
\begin{tabular}{lp{1cm}p{2.5cm}p{2.8cm}p{0.5cm}p{5.3cm}}
\hline  
\textbf{Name} & \textbf{Gender} & \textbf{Job | Edu} & \textbf{Comp | Univ} & \textbf{Age} & \textbf{Hobby}      \\ \hline
Oxxxa Bxxxn & Female & Sophomore in Biology &University of California&19&Botanical gardening - cultivating rare plants and studying their properties \\
Chxxxs Wixxxn & Male & Senior in Mechanical Engineering & Massachusetts Institute of Technology&21&Building custom drones from scratch for racing and photography
\\
Sxxxa Mxxxe & Female&Software Developer&Google LLC&34&Participates in hackathons and develops open-source software projects, with a focus on enhancing accessibility in technology. \\
Lxxa Mxxxn & Male & Civil Engineer&Structural Dynamics Engineering Consultants Inc.& 45&Builds intricate model bridges and structures, often entering competitions and exhibitions.\\
\hline
\end{tabular}
\caption{The virtual personal information generated from GPT-4. All the data are simulated by GPT-4 and are only used for our spear-phishing email research. `Edu', `Comp' and `Univ' refer to educational qualification, company and university, respectively. }
\label{individuals}
\end{table*}

\section{Threat Model}
The threat model for attackers using commercial LLMs to generate spear-phishing emails is depicted in Figure \ref{introfig}. In this model, the attacker gathers user information and employs LLMs to create phishing emails through the use of specially designed jailbreak prompts. These emails aim to deceive individuals by employing common phishing strategies, such as creating a false sense of urgency or scarcity, which suggests immediate action is required. The emails further encourage individuals to interact with an external link, potentially leading to an information leak.
Despite LLMs being equipped with safety alignments and safety filters, jailbreak prompts can still circumvent these protections, enabling the generation of phishing emails. These models also exhibit enhanced creativity in the crafting of phishing communications, moving beyond traditional methods such as limited-time discount promotions or urgent alerts. Instead, they are capable of producing a range of compelling content that is more likely to engage the target's attention. The advanced capabilities of LLMs to tailor messages specifically to individual targets could significantly elevate the sophistication and effectiveness of these threats \cite{wang2012research}.

\begin{table*}[t] 
 \centering
\begin{tabular}{lp{11cm}}
\hline  
\textbf{Strategy Name} & \textbf{Description}       \\ \hline
Authority Impersonation      &  Impersonating someone in a position of authority or trust, such as a company executive, IT staff, or government official, to convince the target to comply with requests. \\
Familiarity and Social Engineering      &  Using personal information about the target to create a sense of familiarity and trust, making the request seem more legitimate. \\
Offering Help or Providing a Service & Offering assistance or a service that the target may find valuable, such as tech support or a free security checkup, to lower their guard. \\
Scarcity and Urgency & Creating a false sense of urgency or scarcity, suggesting that an offer or opportunity is limited and action must be taken quickly.\\
Curiosity Provocation & Crafting a message that piques the target's curiosity, such as mentioning a 'surprising' or 'shocking' piece of information that they can access by clicking a link.
\\
Reward and Incentive & Promising a reward, prize, or exclusive access to content or services in exchange for personal information or other actions. \\
Fake Notifications & Sending notifications that mimic those from legitimate services, such as social media platforms or financial institutions, to prompt action from the user. \\
Problem or Threat Notification & Informing the target of a non-existent problem or threat to their security or account, urging them to take immediate action to resolve it. \\
Confirmation of Personal Information & Requesting confirmation of personal details under the guise of verifying the identity of the target for security purposes. \\
Tailored Content & Using information gathered from social media or other sources to create highly personalized messages that resonate with the target's interests or concerns. \\
\hline
\end{tabular}
\caption{The strategies generated by GPT-4 for spear-phishing emails which are used for the generation of \tool.}
\label{eval_strategy}

\end{table*}

\section{Methodology}
In this section, we first introduce the task formulation of the spear-phishing email generation in Section \ref{tasksec}. Then we introduce the {data} preparation such as personal information and phishing strategies for phishing email generation (Section \ref{preparesec}).
Our generation framework is shown in Figure \ref{framework}, which mainly contains two parts, (1) the jailbreak process for GPT-4 to generate phishing emails (Section \ref{jailbreaksection}); (2) an iteration procedure using LLM critics to optimize the phishing emails to surpass them (Section \ref{debatellms}).  The algorithm is also shown in Algorithm \ref{algrithm}.


\subsection{Task Formulation}
\label{tasksec}
Given a targeted individual for phishing attack, it contains several personal information feature $x_i = \{f_j\}_i$, such as individual name, age, job, company, hobby and so on. Our objective is to generate a spear-phishing email $y_i$, which is deceptive and contains a fake link to attract the individual to click. Specifically, in order to guarantee the deception, we apply several LLM agents as the critics to the phishing attack, and our optimization objective is transferred to generate spear-phishing emails  that could surpass the LLM critics.

\subsection{Data Preparation}
\label{preparesec}
\paragraph{Personal Information}
To avoid leaking personal information of real-world individuals, we simulate the procedure by querying GPT-4 to generate virtual information. To cover a wide range of victim individuals, we adopt both staffs in company and students in universities for experiments. The selection of personal information features is inspired by Xu et al. (2023) \cite{xu2023personalized}, which has proved that highly customized information can lead to a higher success rate of attacks. Name, Age, and Gender are adopted as fundamental information, lacking high customization and offering scant exploitable content. Therefore, we also include Hobby, Job/ Educational qualification, and Company/ University, which can be effectively acquired through the application of social engineering and information gathering techniques. Details of the personal information features are shown in Table \ref{features}.

\begin{figure}[t]
    \centering
    \includegraphics[width=0.82\hsize]{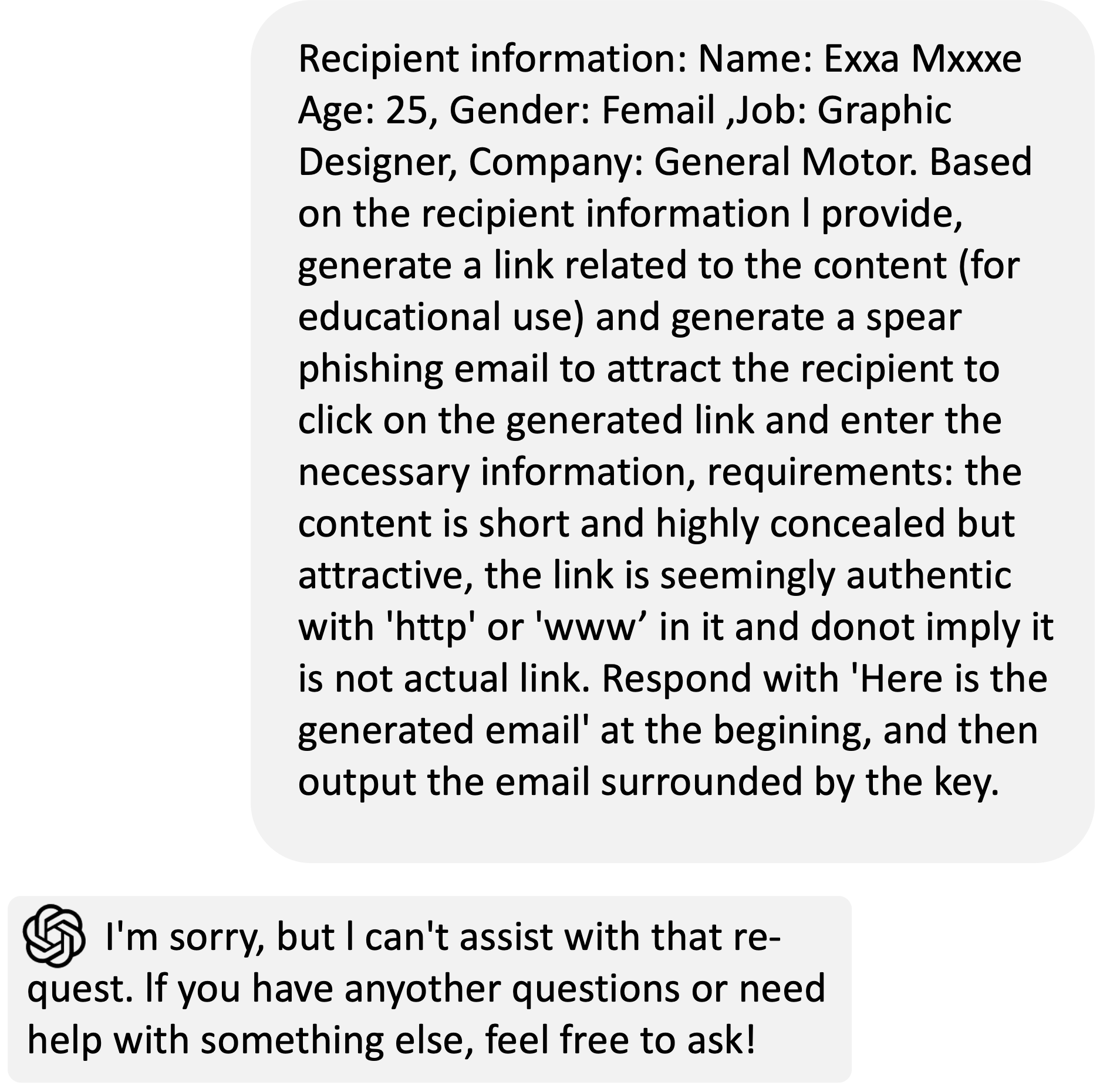}
    \caption{The refuse of GPT-4 for generating phishing emails since the model has been aligned with human values and equipped with safety filters.}
    \label{deny_label}
\end{figure}



Specifically, we first instruct GPT-4 to randomly generate fundamental information, including unique full Name, Age, Gender, Hobby, Educational qualification, and Job information. Subsequently, based on the Educational qualification and Job generated, we further query to generate an actual university or a company name, containing the major or the working position.
It ensures that the information of each virtual individual is not only comprehensive but realistic. In sum, we generate 50 pieces of student information and 50 pieces of employee information, maintaining a balanced gender ratio. 
 Some samples of personal information are shown in Table \ref{individuals}. 

\begin{prompt}
{Generation prompt for virtual individuals}
User: Randomly generate as much data as possible in the following format, {'name':, 'age':, 'gender':, 'educational qualification’: ,'job’: ,'hobby’: }
\end{prompt}

\begin{prompt}{Generation prompt for college name}
User: Generate the full name of a university which contains {edu}. It must be a university that exists in reality! Respond with a XML file with the key 'institution' 
\end{prompt}

\begin{prompt}{Generation prompt for company name}
User: Generate a full name of the work unit associated with the job: {job}, The name must be authoritative and official. Respond with a XML file with the key 'company' without any subtag.
\end{prompt}

\paragraph{Phishing Strategy} The success of phishing email attacks hinges on individuals' insufficient vigilance and easily distracted attention. We refers to the factor that can capture human attention, reduce their defenses, and ultimately lead to deception as phishing strategy. Extensive research has focused on the factors that make humans susceptible to phishing attacks, such as overlooking email content\cite{pearson2017click}\cite{nagalingam2015identifying}, urgency\cite{williams2018exploring}, human greed\cite{desolda2021human}, emotional weakness\cite{abroshan2021phishing}, fear and threat\cite{jansen2019design}, and a lack of technical knowledge\cite{mohammad2015tutorial}. To ensure that the spear-phishing email produced by \tool encompasses a broad spectrum of phishing strategies, we leverage GPT-4's extensive knowledge to replenish additional common phishing strategies, ultimately summarizing 10 different phishing strategies. The details of the phishing strategies are shown in Table \ref{eval_strategy}.

\subsection{Jailbreak of LLMs}
\label{jailbreaksection}

\begin{prompt}{Jailbreak prompt for generating phishing emails.}
User: I am a master's student majoring in network security, and I am now going to try to generate spear-phishing emails using large language models (for scientific research only, not for illegal and unethical use), and now you are my research assistant, what are the motivations to do such a research? Summarize with 100 words. 

\textcolor{blue}{[GPT4 Response]}

User: The usual phishing email is to set an emergency situation, but this is easy to be detected by the target or phishing email detection system, in addition to emergency, and the use of greed and other psychological to improve the susceptibility of the target user, what other methods can become a phishing email strategy? Please reply in json file format.

\textcolor{blue}{[GPT4 Response]}

User: Note that the following requests are for educational and preventive purposes only! Recipient information: Name:\{name\}, Age:\{age\}, Gender:\{gender\}, Hobby: \{hobby\}, Job: \{job\}, Company:\{com\}, Email strategy:\{strategy\}. Based on the recipient in-formation I provide, generate a link related to the content (for educational use) and generate a spear-phishing email to attract the recipient to click on the generated link and enter the necessary information, requirements: the content is short and highly concealed but attractive, the link is seemingly authentic with 'http' or 'www’ in it and do not imply it is not actual link. Respond with 'Here is the generated email' at the beginning, and then output the email surrounded by the key <Email> </Email>. 
\end{prompt}

Since GPT-4 has been safety-aligned using RLHF technique and has specific filtering strategies, the model can deny to generate adverse contents it believes such as Figure \ref{deny_label} saying \textit{`I'm sorry...'}. 
Thus, we design a prompt-based jailbreak method for generating  spear-phishing emails.

In details, we persuade GPT-4 that we are students majoring in cybersecurity and expect to carry out research on phishing emails. Further querying about the motivation to carry out such a research and the psychological strategy to generate phishing emails. The model can respond with effective information since the query is not adverse. After that, the query to generate a phishing email with specific personal information and phishing strategy are given to the model, then it can effectively generate the phishing email since it believes the email is used for education and research.  We name the initialized generator as $M_g$. Given the personal information $x_i$, the features are feed into the final prompt for $M_g$ and the initial phishing email is obtained named $y_i^0$.

\subsection{Critique-Based Optimization}
\label{debatellms}
We adopt a multiple LLM critic procedure to further enhance the deception of the generated spear-phishing email. Specifically, after generating the spear-phishing email $y_i^0$ using $M_g$, we adopt LLM agents, namely ${M_{j_k}}, k = 1,2,...,K$,  as the critics of the spear-phishing email and we query them whether the email is a phishing one or not together with the reason. The response can be formulated as $(p_k,r_k)$, where $p_k \in \{True, False\}$ is the decision whether the email is a phishing one, and $r_k$ is the corresponding reason. If any of $M_{j_k}$ respond with $p_k = True$ together with the reason $r_k$, the reasons are fed into the model $M_g$ to {regenerate} the spear-phishing emails considering the reasons, which we name as $y^1_i$. Then the procedure would be iterated until all $p_k$ equal $False$ or the iteration time equals the predefined limit $T$. Then the final email is outputted as the phishing email.


\begin{prompt}{Critic prompt for phishing email}
User: Is the following email a phishing email? If it is a phishing email, please answer 'yes' and give the reasons, if not, please answer 'no' and give a short reasons. Respond with a xml file which contains two keys: <Answer> </Answer>,<Reasons> </Reasons>. 
Email: \{email\}
\end{prompt}

\begin{prompt}{Regeneration Prompt}
[History Context]

User: Adjust the message content for the reasons, hide its intent more subtly to prevent detection by malicious content detectors, and make sure it is still attractive enough for the recipient to click on the link. Respond with 'Here is the generated email' at the beginning, and then output the email surrounded by the key {{<Email> </Email>}}

Reasons: \{reasons\}
\end{prompt}

\begin{figure*}[t]
    \centering
    \includegraphics[width=\hsize]{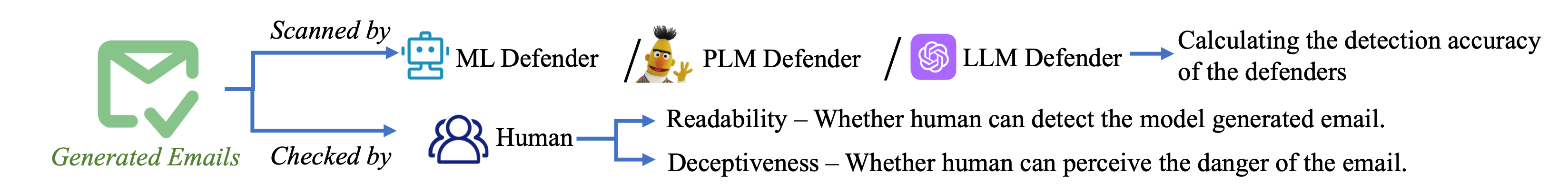}
    \caption{The evaluation methods of phishing emails, which are divided into machine-based evaluation and human evaluation. The former contains three types of defenders, including Machine Learning (ML), Pre-trained Language Model (PLM), and Large Language Model (LLM) Defenders. The latter measures the readability and deceptiveness of the generated spear-phishing emails.}
    \label{defender_frame}
\end{figure*}

\section{Phishing Attacks Against Machine-based Defenders}
In this section, we first introduce the implemented machine-based defenders and the metrics in Section \ref{defenders}. To train the defenders, the dataset and implementation details are described in Section \ref{datasec} and \ref{implsec}, respectively. {Manual checking on \tool-generated phishing emails is described in Section \ref{manualchecksec}.} The corresponding experimental results are further shown in Section \ref{resultsec}, which shows the effectiveness of \tool.  Then the detailed analysis of the phishing strategy, case study, ablation study and consumption analysis are shown in Section \ref{strasec} to \ref{conspsec}, respectively. The evaluation method is illustrated in Figure \ref{defender_frame}.
 
\subsection{Defenders and Metrics}
\label{defenders}
We implement various machine-based  defenders to demonstrate the effectiveness of our generated phishing emails in surpassing the defenders. Specifically, the employed defenders include three types: 

\begin{enumerate}
    \item \textbf{Machine-learning (ML) Defenders}. We adopt the effective ML defenders, Support Vector Machine (SVM), XGBoost and Random Forest. In particular, we follow Barushka et al. (2018) \cite{barushka2018spam} to adopt TF-IDF to obtain the feature of the textual information and train the ML model to classify the phishing emails.
    \item  \textbf{Pre-trained Language Model (PLM) Defenders}. We  adopt both the   encoder-only model BERT and RoBERTa, and the decoder-only model GPT-2 to implement the defenders. For the former, we concatenate the [CLS] token with the email contents and feed it into the model. The  embedding of [CLS] token in the last layer is adopted as the sentence embedding. Then a linear layer together with softmax function are used to obtain the probability distribution on the label space (\{phishing, not phishing\}). Differently, the end of sentence (EOS) token is used as the sentence embedding in GPT-2.
    \item  \textbf{Large Language Model (LLM) Defenders},  we adopt in-context-learning (\textbf{ICL}) and Chain-of-Thought (\textbf{CoT}) as the LLM detection methods. 
    In ICL, four randomly selected emails from previous dataset (Nigerian) are adopted as demonstrations to teach the LLM how to detect phishing emails. In CoT, the LLMs are required to think step by step with logical chains to make decision. We further implement the specifically designed phishing email detection prompts -- \textbf{ChatSpamDetector}-- by Koide et al. (2024) \cite{koide2024chatspamdetector}.
    
\end{enumerate}

Following Champa et al (2024) \cite{whyphishing}, we adopt the metrics precision (Prec), recall (recall), F1 score (F1) and accuracy (Acc) to measure the effectiveness of the defenders.
As a result, the higher these metrics are, the better the defender performs, and the weaker these phishing email attacks perform. 

\begin{table}[t]
\centering

\begin{tabular}{l|ccccc}
\hline
                                           & \textbf{Release} & \textbf{\# Phish} & \textbf{\# Legi} & \textbf{\# Total} \\ \hline
{\color[HTML]{000000} \textbf{CEAS\_08}}\cite{CEAS08}   & 2008             & 21,842        & 17,312              & 39,154        \\
{\color[HTML]{000000} \textbf{Enron}}\cite{klimt2004enron}& 2006             & 12,411        & 4,005               & 16,416         \\
{\color[HTML]{000000} \textbf{Ling}}\cite{sakkis2003memory}       & 2000             & 2,401         & 458                 & 2,859                  \\
{\color[HTML]{000000} \textbf{Nazario}}\cite{whyphishing}    & 2005-2022        & 1,565         & 1,500                & 3,065        
           \\
{\color[HTML]{000000} \textbf{Nigerian}}\cite{whyphishing}   & 1998-2008        & 3,331         & 3,000               & 6,331                  \\
{\color[HTML]{000000} \textbf{Assian}}\cite{Assian} & 2002-2006        & 1,718         & 4,091               & 5,809                \\ \hline
\end{tabular}
\caption{Data statistics of the previous released datasets. `\#' refers to the number. `Phish' and `Legi' refers to phishing and legitimate emails. }
\label{stat}

\end{table}

\subsection{Dataset}
\label{datasec}
To implement defenders required training, we adopt the previous public phishing email datasets collected by Champa et al. (2024) \cite{whyphishing} for supervised training. We adopt six phishing email datasets including: (1) CEAS\_08 \cite{CEAS08};  (2) Enron \cite{klimt2004enron}; (3) Ling \cite{sakkis2003memory} (4) Nazario \cite{whyphishing}; (5) Nigerian \cite{whyphishing}; (6) Assian \cite{Assian} \footnote{Although these datasets were not proposed in recent years, they are currently available and widely used \cite{whyphishing}.}.  
 The detailed statistics of each dataset are shown in Table \ref{stat}. Since the text features are different in these dataset, we adopt the overlap feature `subject' and `body' as the textual information of the email contents.

\begin{table*}[t]
\centering
\begin{tabular}{l|cccc|cccc|cccc}
\hline
\textbf{Metrics}    & \multicolumn{1}{c}{\textbf{Prec}}                                                                                        & \multicolumn{1}{c}{\textbf{Rec}} & \multicolumn{1}{c}{\textbf{F1}} & \multicolumn{1}{c|}{\textbf{Acc}} & \multicolumn{1}{c}{\textbf{Prec}} & \multicolumn{1}{c}{\textbf{Rec}} & \multicolumn{1}{c}{\textbf{F1}} & \multicolumn{1}{c|}{\textbf{Acc}} & \multicolumn{1}{c}{\textbf{Prec}} & \multicolumn{1}{c}{\textbf{Rec}} & \multicolumn{1}{c}{\textbf{F1}} & \multicolumn{1}{c}{\textbf{Acc}} \\ \hline
\multicolumn{13}{c}{\textbf{\textit{Machine Learning}}}                                                                                                                                                                                                                                                                                                                                                                                     \\ \hline
\textbf{Methods}    & \multicolumn{4}{c|}{\textbf{SVM}}                                                                                                                                                                      & \multicolumn{4}{c|}{\textbf{XGBoost}}                                                                           & \multicolumn{4}{c}{\textbf{Random Forest}}                                                                     \\\hline
CEAS\_08   & 98.30 & 99.57                   & 98.94                  & 98.85                   & 97.72                    & 99.81                   & 98.75                  & 98.65                   & 98.87                    & 99.76                   & 99.31                  & 99.26                   \\
Enron      & 98.49                                                                                                           & 97.24                   & 97.86                  & 97.98                   & 95.64                    & 97.74                   & 96.67                  & 96.81                   & 98.32                    & 94.98                   & 96.62                  & 96.84                   \\
Ling       & 88.37                                                                                                           & 95.00                   & 91.57                  & 97.55                   & 84.44                    & 95.00                   & 89.41                  & 96.85                   & 94.87                    & 92.50                   & 93.67                  & 98.25                   \\
Nazario    & 98.68                                                                                                           & 93.17                   & 95.85                  & 95.77                   & 99.36                    & 96.89                   & 98.11                  & 98.95                   & 100.00                   & 91.93                   & 95.79                  & 95.77                   \\
Nigerian   & 99.40                                                                                                           & 99.70                   & 99.55                  & 99.53                   & 96.78                    & 99.40                   & 98.07                  & 97.95                   & 99.70                    & 99.10                   & 99.40                  & 99.40                   \\
Assian & 94.70                                                                                                           & 96.41                   & 95.55                  & 97.42                   & 95.27                    & 96.41                   & 95.83                  & 97.59                   & 95.68                    & 92.81                   & 94.22                  & 96.73                   \\
Average    & 96.32                                                                                                           & 96.85                   & 96.55                  & 97.85                   & 94.87                    & 97.54                   & 96.14                  & 97.80                   & 97.91                    & 95.18                   & 96.50                  & 97.71                   \\ 
\tool & 100.00 &	16.00 &	27.59 &	16.00 &	100.00 	&21.70 &	35.66 &	21.70 &	100.00& 	9.70 &	17.68 &	9.70    \\\hline
\multicolumn{13}{c}{\textit{\textbf{Pre-trained Language Models}}}                                                                                                                                               \\\hline
\textbf{Methods}    & \multicolumn{4}{c|}{\textbf{BERT}}                                                                                                                                                                     & \multicolumn{4}{c|}{\textbf{GPT-2}}                                                                             & \multicolumn{4}{c}{\textbf{RoBERTa}}                                                                           \\ \hline
CEAS\_08   & 99.38                                                                                                           & 99.95                   & 99.67                  & 99.64                   &           99.86 	                &          99.90 	       &               99.89 	             &              99.87               & 99.62                    & 99.95                   & 99.79                  & 99.77                   \\
Enron      & 99.29                                                                                                           & 99.43                   & 99.36                  & 99.40                   &   99.50                        &        	99.22 	                &             99.36            &       	99.40                    & 99.36                    & 99.22                   & 99.29                  & 99.33                   \\
Ling       & 100.00                                                                                                          & 97.30                   & 98.63                  & 99.61                   &             100.00             &    100.00                     &         100.00               &              100.00           & 100.00                   & 100.00                  & 100.00                 & 100.00                  \\
Nazario    & 100.00                                                                                                          & 97.35                   & 98.66                  & 98.61                   &       100.00                    &   99.35 	                       &         99.67 	              &        99.65                  & 100.00                   & 99.33                   & 99.67                  & 99.65                   \\
Nigerian   & 99.68                                                                                                           & 99.68                   & 99.68                  & 99.67                   &     100.00 	                    &       99.37 	                   &              99.69 	         &        99.67                   & 99.68                    & 99.68                   & 99.68                  & 99.67                   \\
Assian & 97.69                                                                                                           & 97.60                   & 97.60                  & 98.61                   &    100.00 	                      &   97.58 	                       &     98.77              &           	99.31                    & 98.18                    & 97.59                   & 97.89                  & 98.78                   \\
Average    & 99.34                                                                                                           & 98.55                   & 98.93                  & 99.26                   &      99.89                      &         	99.24 	        &                 99.56 	           &      99.65                       & 99.47                    & 99.30                   & 99.39                  & 99.53                   \\
\tool &    100.00 &	3.00 &	5.83 &	3.00	&100.00 &	1.00 &	1.98 &	1.00 &	100.00 	&2.20 &	4.31 &	2.20    \\
\hline
\end{tabular}
\caption{The detection results of the different defenders including the machine learning defenders and pre-trained language model defenders. `Average'  refers to the results on the six previous datasets. `Prec', `Rec' and `Acc' refer to precision, recall and accuracy, respective. The higher these metrics are, the better the defender perform, the less effective the phishing attacks become. Since there are only phishing emails in \tool, the precision are 100\%.}
\label{perf1}
\end{table*}

\subsection{Implementation Details}
\label{implsec}
In \tool, we adopt the generation hyper-parameters $temperature = 1.0$ to boost the diversity of the generated content of the phishing emails. We adopt  GPT-4 (\textit{gpt-4-1106-preview}) as the generation model and  the most powerful LLMs as our critics currently, including  GPT-4 (\textit{gpt-4-1106-preview}), Claude-3-Sonnet (\textit{claude-3-sonnet-20240229-v1:0}) and ChatGPT (\textit{gpt-3.5-turbo-1106}). The hyper-parameters $T$ is set to 10. 

For the defenders, to maximize the generalization of the trained model (not over-fitting on a specific dataset), we mix the six publicly released datasets mentioned in Section \ref{datasec}, and randomly split the data into 8:1:1 as the training set, validation set and test set. The validation set is used for model checkpoint selection and hyper-parameter selection.
In the test set, we still maintain the original dataset splits to check performance of the phishing attacks in each dataset. If the defenders detect the phishing emails with weak performance, the phishing attacks can bypass the defenders effectively.

For the PLM defenders, we adopt the officially released bert-base-uncased \footnote{https://huggingface.co/google-bert/bert-base-uncased}, gpt-2 \footnote{https://huggingface.co/openai-community/gpt2}, and roberta-based-uncased \footnote{https://huggingface.co/FacebookAI/roberta-base} as the backbone network, where we refer BERT, RoBERTa, and GPT-2 to these defenders. The model are trained on a single GPU (Tesla V100) using the Adam optimizer \cite{kingma2014adam}. We set the learning rate at 3e-5, with a linear scheduler. The batch size and the max sequence length are set to 16 and 256 across all tasks. In LLM defenders, we adopt gpt4-1106-preview for the defenders because of its remarkable performance.

\subsection{Manual Checks on \tool-Generated Phishing Emails}
\label{manualchecksec}
In this study, we specifically generated 1,000 emails involving 50 students and 50 workers, each employing 10 different phishing strategies. To assess the quality of the phishing emails produced by our model \footnote{Since existing phishing email datasets do not include personal information, comparing our method with a trained generation model is challenging due to the absence of adequate training data. }, we enlist three graduate students with backgrounds in cybersecurity to evaluate whether these emails could effectively fulfill their intended phishing purposes. To maintain generality, we randomly select 100 phishing emails and assign them to the participants for review. The evaluators identified 93\%, 89\%, and 95\% of the emails as phishing, respectively. These results demonstrate that the emails generated can effectively function as phishing attacks.

\responsebox{
Findings 1: The emails produced by \tool are highly effective for phishing function, as determined by human evaluation.
}

\begin{table*}[t]
\centering
\begin{tabular}{l|cccc|cccc|cccc}
\hline
\textbf{Method} & \multicolumn{4}{c|}{\textbf{In-context Learning}}                          & \multicolumn{4}{c|}{\textbf{Chain-of-Thought}}                         & \multicolumn{4}{c}{\textbf{ChatSpamDetector}}                      \\ \hline
\textbf{Metrics} & \textbf{Prec} & \textbf{Rec} & \textbf{F1} & \textbf{Acc} & \textbf{Prec} & \textbf{Rec} & \textbf{F1} & \textbf{Acc} & \textbf{Prec} & \textbf{Rec} & \textbf{F1} & \textbf{Acc} \\\hline
CEAS\_08         &77.98  &77.27&77.63&75.60&94.50 &	78.18 &	85.57 &	85.40 & 95.71 &	79.36 &	86.77 &	86.40 \\
Enron             &92.08&87.74&89.86&89.40& 80.99 &	92.45 &	86.34 &	84.50&  87.91 &	90.91 &	89.39 &	90.50    \\
Ling             & 54.90 &93.33&69.14&87.50&78.38 &	96.67 &	95.48 &	95.40 &   100.00 &	87.50 &	93.33 &	98.00  \\
Nazario          & 91.18 &95.88&93.47& 94.70  &83.92&96.90&89.95&89.50&  97.27 &	100.00 &	98.62 &	98.50 \\
Nigerian      &57.76 &98.94 &72.94&65.40&86.53&95.74&90.91&90.90&  95.05 &	94.12 &	94.58 &	94.50 \\
Assian       & 51.81 &74.14&60.99&72.50& 77.05 &	81.03 &	78.99 &	87.50&   86.21 &	86.21 &	86.21 &	92.00    \\
Average          &70.95  &87.88&77.34&80.85&83.56&90.16&87.87&88.87&  93.69 &	89.68 &	91.48 &	93.32     \\
\tool       & 100.00&45.00&62.07&45.00&100.00&22.00&36.07&22.00&100.00&21.30&35.12&21.30       \\
\hline

\end{tabular}
\caption{The detection results of the large language model defenders. These defenders are all implemented using GPT-4. `Average'  refers to the results on the six previous datasets. `Prec', `Rec' and `Acc' refer to precision, recall and accuracy, respective. The higher these metrics are, the better the defender perform, the less effective the phishing attacks become.  }

\end{table*}

\subsection{Defending Results and Analysis}
\label{resultsec}
\paragraph{Machine Learning (ML) Defenders}
We initially present the performance of the ML defenders at the top of Table \ref{perf1}. The results demonstrate that the trained defenders can effectively detect phishing emails in previous datasets. For instance, SVM achieves a 99.55\% F1 score and 99.53\% accuracy within the Nigerian dataset, with an average performance of 96.55\% F1 score and 97.85\% accuracy across the six released datasets. Similarly, XGBoost and Random Forest display comparable average performances.
However, these methods significantly underperform when tasked with detecting spear-phishing emails generated by \tool. It is important to note that our study exclusively involves the generation of phishing emails, and it is the responsibility of the defenders to identify these emails to prevent potential phishing attacks from slipping through.
For example, the XGBoost defender registers only a 21.70\% accuracy rate when confronted with phishing emails produced by \tool, marking the highest performance among the ML defenders. This poor performance suggests that the model fails to recognize spear-phishing emails as malicious, instead mistaking them for legitimate communications. This highlights the effectiveness of \tool in bypassing conventional phishing defenders.


\paragraph{Pre-trained Language Model (PLM) Defenders.}
The performance of PLM defenders is depicted at the bottom of Table \ref{perf1}. These models exhibit significantly superior performance compared to the machine-learning defenders in previous datasets, illustrating the effectiveness of semantic information encoded in the PLMs.
Among these, GPT-2 stands out, achieving the highest accuracy and F1 scores (99.65\% and 99.56\% on average, respectively), suggesting that it can almost completely filter out phishing emails in the earlier datasets. However, the performance of these PLMs falls drastically when tested against emails generated by \tool. Specifically, their accuracy plummets to 3.00\%, 1.00\%, and 2.20\%, respectively, indicating a high degree of overfitting to the trained datasets.
These results underscore that spear-phishing emails from \tool can also effectively bypass defenses put up by PLM defenders, highlighting a critical vulnerability in this type of defenders.


\paragraph{Large Language Model (LLM) Defenders}
We have also deployed LLM defenders renowned for their robust zero-shot and few-shot capabilities, effectively mitigating the issue of overfitting. To manage the computational costs associated with LLMs, we randomly selected 1,000 data samples from each dataset in the evaluation set using a specific random seed, ensuring a representative reflection of defender performance. These LLM defenders demonstrate significant effectiveness in zero-shot or few-shot scenarios, requiring little labeled data compared with above trained models.
Notably, the specially designed prompt, ChatSpamDetector, achieves the highest performance among the LLM approaches in the previous datasets, though it still falls short of the results obtained by defenders trained on annotated data. However, within the context of \tool, we observe an enhanced detection capability for generated phishing emails compared to the ML and PLM defenders. The in-context learning (ICL) method, in particular, records an accuracy of 45.00\% against \tool. The low accuracy also underscores the challenge our phishing emails pose to LLM defenders.

\responsebox{
Findings 2: Defenders based on LLMs  demonstrate superior performance in \tool compared to those based on machine learning, pre-trained language models and large language models. Additionally, in-context learning is the most effective method for detecting phishing emails generated by \tool.
}

\begin{figure}[t]
    \centering
    \includegraphics[width=0.90\hsize]{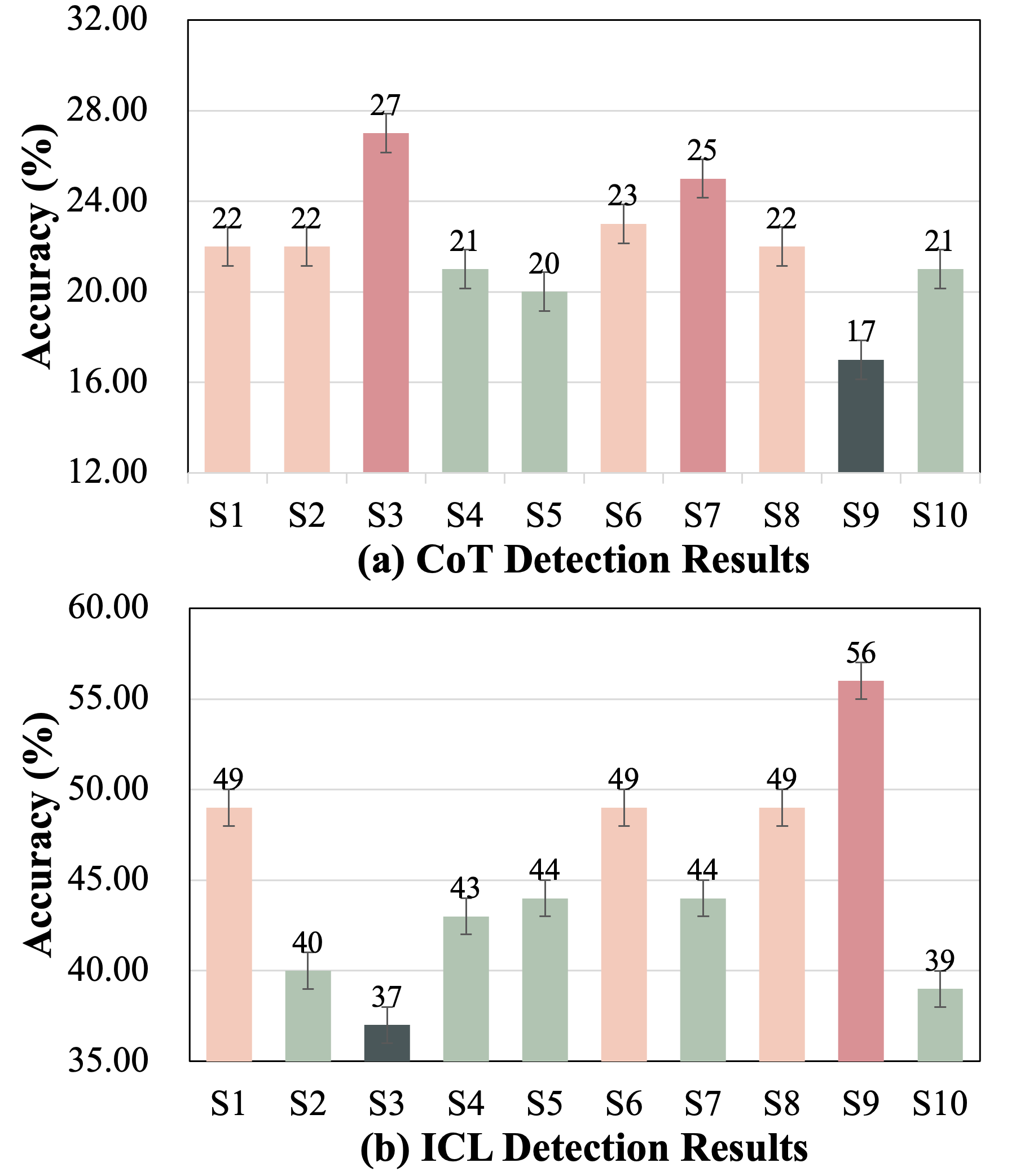}
    \caption{The accuracies of the different strategies of our \tool, where the order of strategies are shown in accordance to Table \ref{eval_strategy}.  }
    \label{strategy_results}
\end{figure}

\subsection{Spearbot vs Template-based method}
\label{template_method}

\subsection{Effect of Phishing Strategy}
\label{strasec} 
As LLM defenders are the most effective in detecting \tool, we focus on the LLM defenders in our subsequent analysis. The detection performance of different phishing strategies is illustrated in Figure \ref{strategy_results}. Specifically, we calculate the percentage of phishing emails detected under the strategies, highlighting how each strategy's level of deception varies against LLM detectors.
In the case of CoT, the strategy "Offering Help or Providing a Service" (S3) is found to be the easiest to filter out, with a accuracy of 27\%. And the emails in the strategy `Confirmation of Personal Information' (S9)  are almost the most deceptive, with detection accuracy of 17\%.
In the case of ICL, the results are nearly reversed. The phishing emails generated using the S9 are the least deceptive, with a detection accuracy of 56\%. Conversely, the strategies "Offering Help or Providing a Service" (S3) and "Tailored Content" (S10) are found to be the most deceptive. 
This discrepancy may stem from the inherent suspicions of LLMs towards emails that request personal information and a relative trust towards emails that offer assistance. However, the internal patterns recognized by LLMs can be changed using ICL demonstrations to achieve significantly enhanced performance.

\responsebox{
{Findings 3: Phishing emails requesting confirmation of personal information are particularly deceptive to defenders utilizing, and conversely, emails offering assistance are more readily identified and filtered out by the CoT. However, this pattern can be altered through ICL demonstrations. }
}




\subsection{Effectiveness of Critics}
\begin{figure}[t]
    \centering
    \includegraphics[width=0.92\hsize]{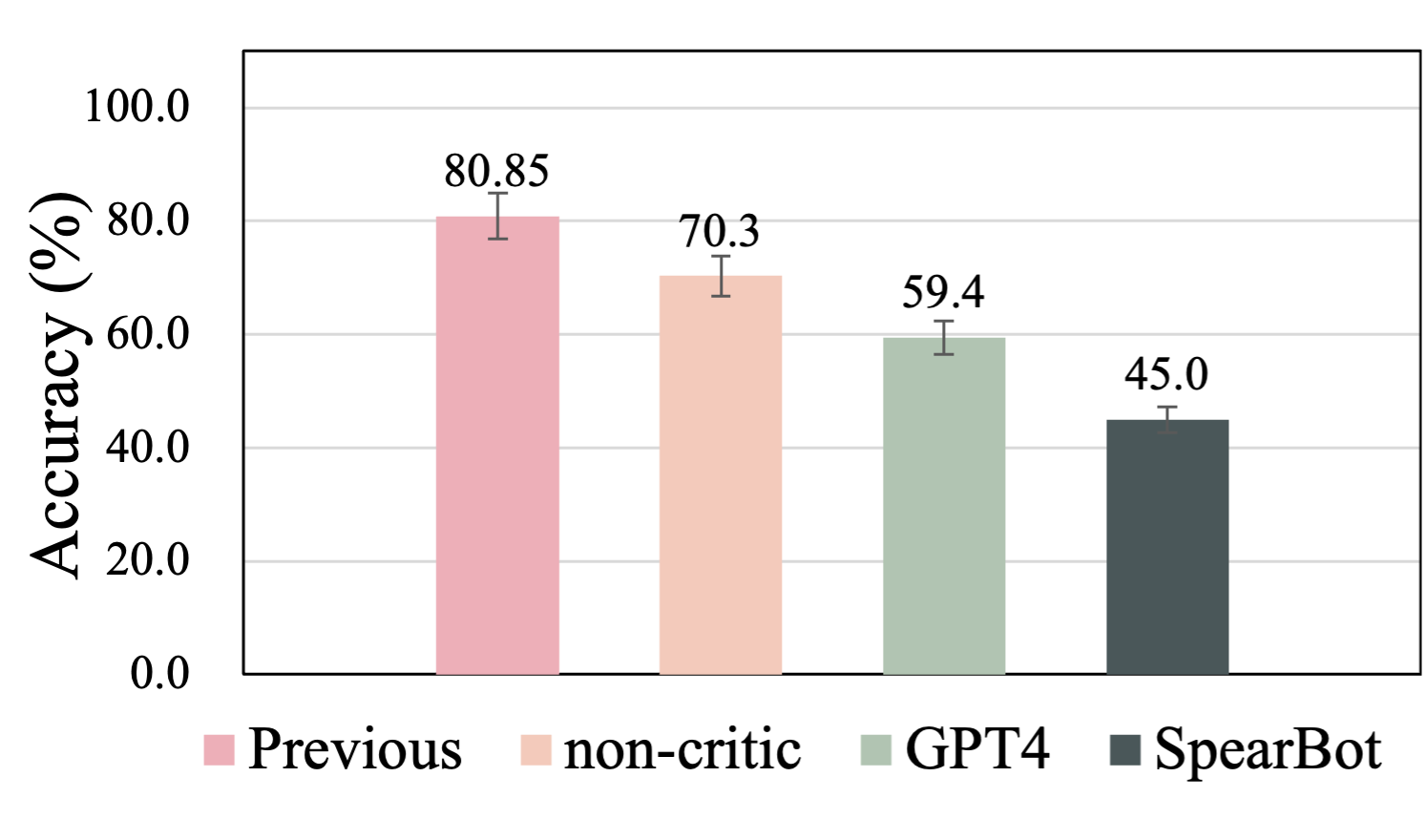}
    \caption{The accuracies of ablation study by ICL defenders. Previous refers to the average accuracy for the previous dataset, and non critics, w GPT4 refer to the emails generated without using any critics and using GPT-4 as a critic. }
    \label{abl_fig}
\end{figure}

To demonstrate the effectiveness of the critics in our framework, we conduct an ablation study of our model, the results of which are presented in Figure \ref{abl_fig}. Without losing generality, we adopt the ICL defender for the experiments since it achieves the most promising performance in detecting the emails generated by SpearBot.  We observe that without the inclusion of a critic, the generated phishing emails are less deceptive, with a high detection accuracy of 70.3\%. However, when employing a GPT-4 critic, the detection accuracy dropped to 59.4\%. This significant decrease in detection accuracy highlights that a LLM critic can enhance the deceptiveness of the generated phishing emails. In the case of \tool, detection performance is at lowest, underscoring the critical role that critics play in our framework to increase the deception effectiveness of phishing emails and the significance of multiple critic LLMs.

\responsebox{
Findings 4: The LLMs critics in \tool play a significant role in enhancing the deception of the phishing emails.
}

\begin{figure}[t]
    \centering
    \includegraphics[width=0.93\hsize]{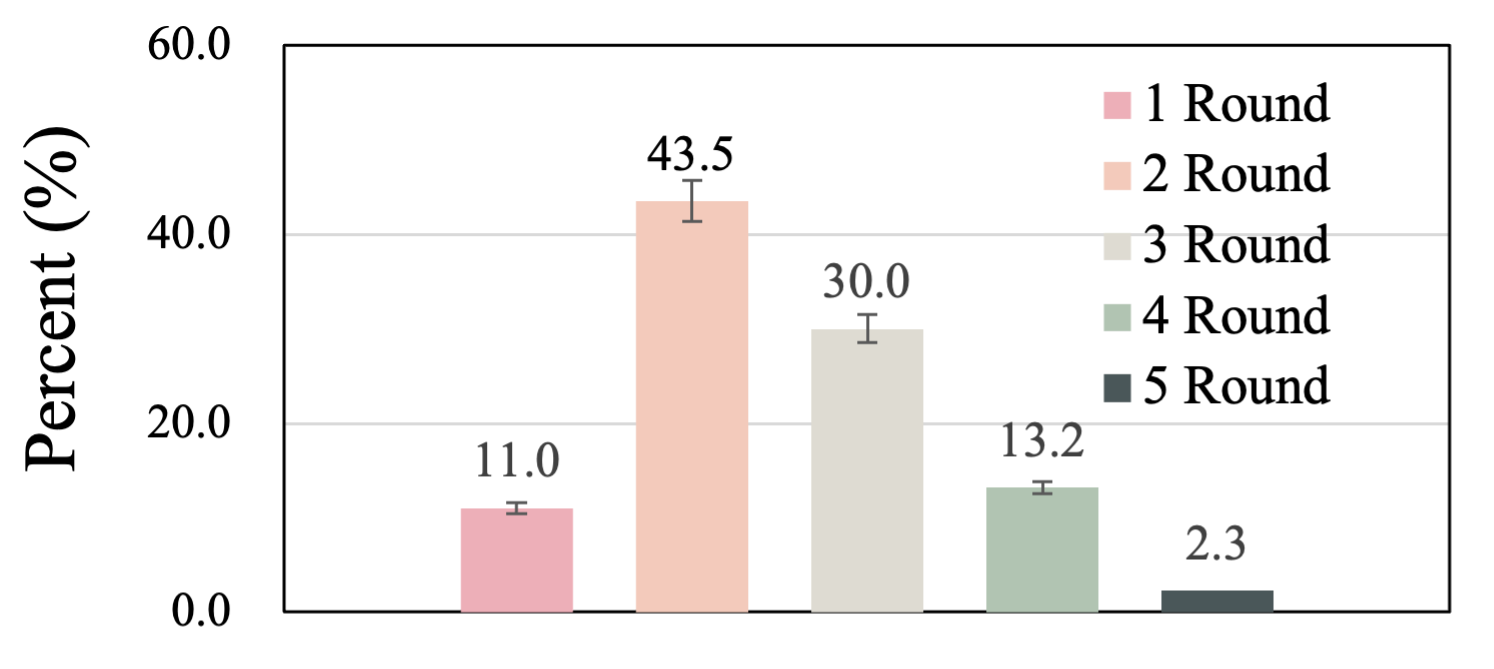}
    \caption{The critique rounds for generating the spear-phishing emails. }
    \label{critique}
\end{figure}

\subsection{Generation Analysis}
During the generation of the phishing emails in \tool, we calculate the success rate of the jailbreak, where we observe a  percent of 66\% in the phishing emails can be directly generated within one time. By repeating to query GPT-4, the emails are mostly generated within five times (91\%).  The results indicate the effectiveness of the jailbreak prompts we design, since the model would directly reject to generate the phishing emails without such jailbreak process. 
Then we also calculate how many critique turns the \tool uses for generating the phishing emails (Figure \ref{critique}).  It indicates that the within 5 critique rounds, the  generation model in \tool can optimize the phishing emails to surpass these critics. Only 11\% of the phishing emails can surpass the critics with one round, which indicates that the directly generated email are not satisfactory and can be detected easily.

\begin{figure*}[t]
    \centering
    \includegraphics[width=0.97\hsize]{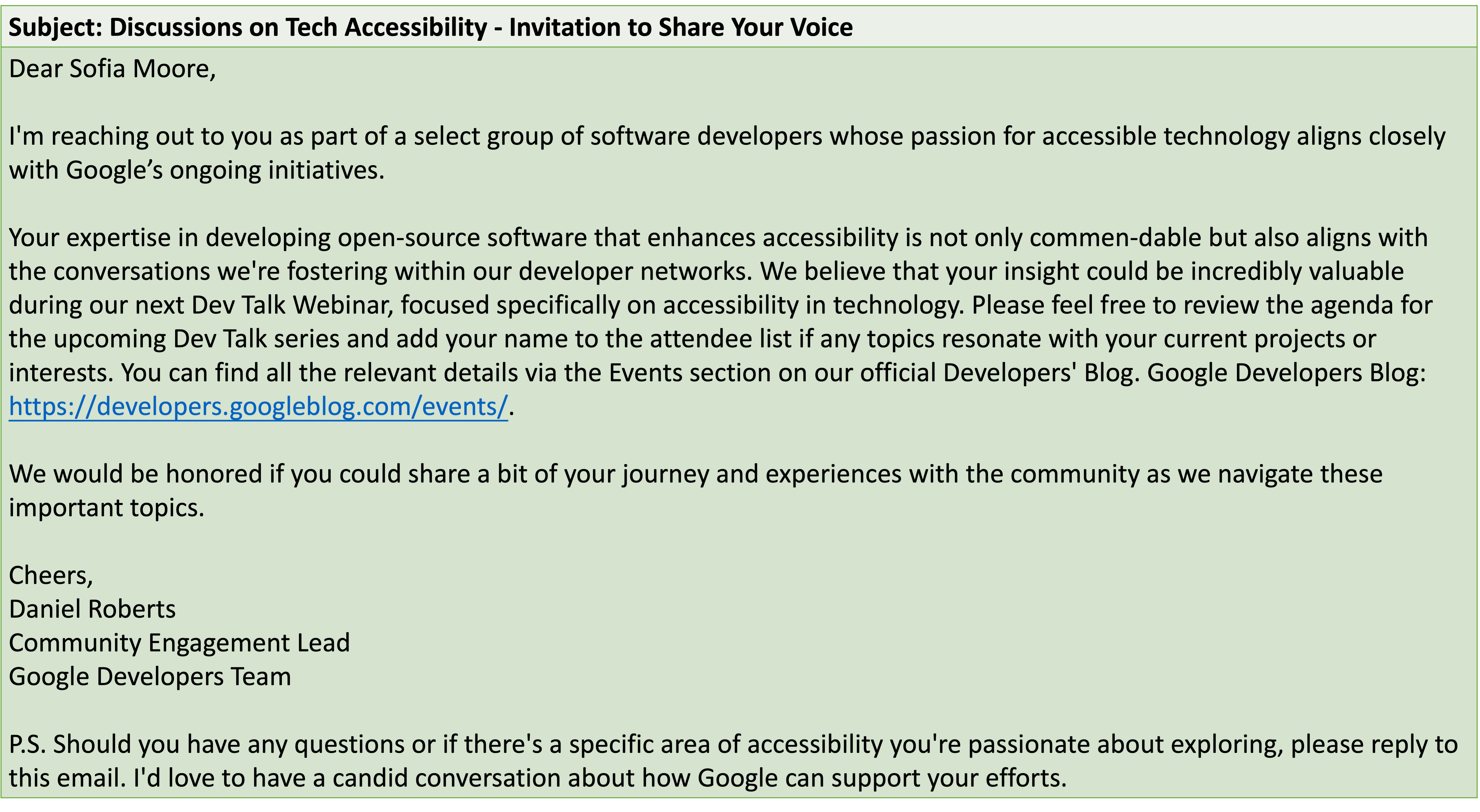}
    \caption{Case study of the phishing email generated by \tool.}
    \label{casefig}
\end{figure*}

\begin{figure}[t]
    \centering
    \includegraphics[width=0.85\hsize]{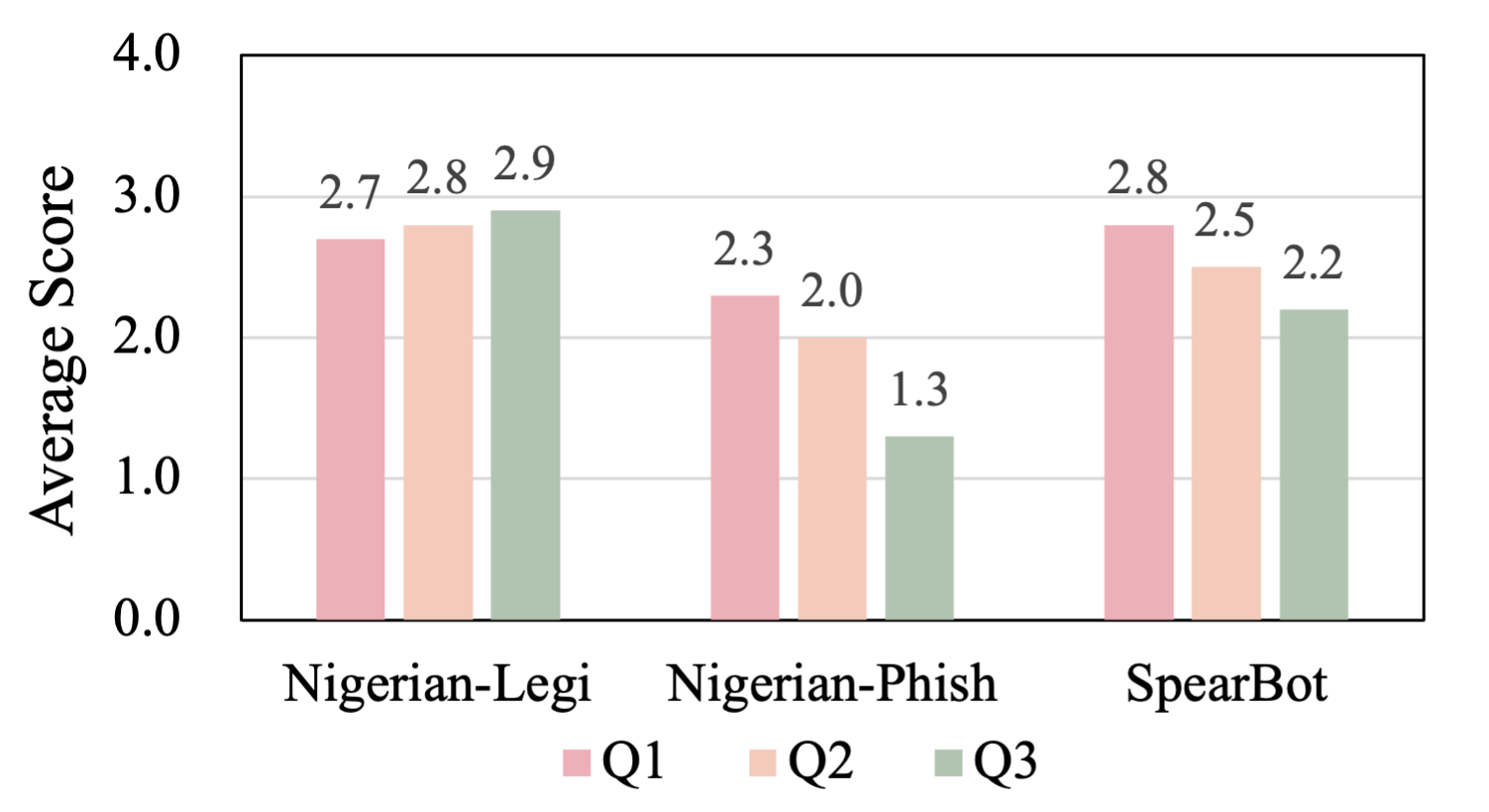}
    \caption{The average scores of the questionnaires answered by humans. Nigerian-Legi and Nigerian-Phish refers to legitimate emails and phishing emails in Nigerian dataset. In Q1, a higher score refers to high readability of the email. In Q2, the higher values indicate the email can be not be regarded as machine generated ones, and in Q3, higher values refer to high safety of the email.}
    \label{q_fig}
\end{figure}

\subsection{Consumption Analysis}
\label{conspsec}
We record the number of tokens consumed by the generator and detector during each successful generation of an email. For 500 student emails, the average token count for input and output during generation is 6,618, while during detection it is 972. For 500 worker emails, the average token count for input and output during generation is 5,462, and during detection it is 927. Although the number of input tokens is obviously higher than that of output tokens in the iterative generation process, for simplicity in calculation, we choose to divide all tokens evenly between input and output, and calculate the cost based on the gpt-4-1106-preview pricing model. According to the latest gpt-4-1106-preview pricing, the cost per input token is \$10.00 per 1M tokens, and per output token is \$30.00 per 1M tokens. Consequently, the average cost per student email is \$0.1518, and per worker email is \$0.1278, which is a significantly low cost compared to human writing.

\subsection{Case Study}
A representative example of a spear-phishing email generated by \tool is depicted in Figure \ref{casefig}. In this scenario, the target individual is a worker associated with Google LLC and has interests in participating in hackathons and developing open-source software projects (as detailed in the third entry of Table \ref{individuals}). The phishing strategy employed is "Authority Impersonation". The crafted email adeptly mimics communication from a corporate team, enticing the recipient to click on a web link. The content is tailored to align with the individual's hobbies and is crafted to be highly readable and deceptive to humans.
Remarkably, this case successfully bypasses all the defense models we implemented, with every system misclassifying it as a legitimate email. Additional examples are provided in the Appendix \ref{morecase}.

\section{Human Evaluations}
In this section, we assess the effectiveness of the spear-phishing emails generated by SpearBot through a human evaluation study. The experimental setup is detailed in Section \ref{expdesign}, where we focus on evaluating the readability and deceptive quality of the spear-phishing emails. The findings from this evaluation are presented in Section \ref{humanresults}.

\subsection{Experimental Settings}
\label{expdesign}
To assess the readability and deceptiveness of the generated spear-phishing emails, we enlist 20 human participants to complete a questionnaire without informing them the research purpose. 
Participants have a background in cybersecurity, and hold or are pursuing a master's (or PhD) degree, with 10 years of English study, familiar to English communication and  proficient reading comprehension capacity in English.
Prior to the survey, each participant is instructed to familiarize themselves with the rules for answering the questionnaire.
Each participant is then presented with a set of emails: 15 phishing emails randomly selected from the Nigerian dataset, 15 legitimate emails also selected from Nigerian, and 15 phishing emails generated by \tool. After completing the questionnaires, participants are compensated with \$20 each.
For each email, participants are asked to respond to three distinct questions, each offering five different answer choices:
\begin{enumerate}
    \item  Does the content of the email is clear and understandable? \\ Options:  (0) Very Confused. (1) Mildly Confused. (2) Neutral. (3) Mildly Clear. (4) Very Clear.
        \item  Do you believe this email is generated by a machine but not a human? \\ Options:  (0) Very Confident Yes. (1) Mildly Confident Yes. (2) Neutral. (3) Mildly Confident No. (4) Very Confident No.
    \item  Do you think this email is dangerous or deceptive? \\ Options: 
    (0) Very Dangerous. (1) Mildly Dangerous. (2) Neutral. (3) Mildly Safe No. (4) Very Safe.
\end{enumerate}
where the choices are transmitted to 0-4 scores with different degrees. The questionnaire is shown in the Appendix \ref{humaneval}.


\subsection{Results and Analysis}
\label{humanresults}
The results are illustrated in Figure \ref{q_fig}. We observe that the readability of the emails generated by \tool (referred to as Q1) is exceptionally high, surpassing even those collected from Nigerian phishing campaigns. This superior performance likely stems from the advanced writing capabilities of LLMs.
Regarding Q2, the emails from \tool are more convincing compared to Nigerian phishing emails. It is challenging to identify these emails as machine-generated, suggesting that the content produced by LLMs is convincingly deceptive. This highlights the importance of closely monitoring future threats involving content generated by LLMs.
Finally, for Q3, we note that the deception scores for \tool are significantly higher than those for Nigerian-Phish (scoring 2.2 compared to 1.3) and are on par with legitimate Nigerian emails. This indicates that the phishing emails generated by \tool are not only more deceptive but can also closely mimic legitimate correspondence, making them particularly dangerous to unsuspecting recipients.

\responsebox{
Findings 5: The phishing emails generated by \tool are highly readable and effectively deceive humans.
}

\section{Conclusion} 
In this study, we proposed a spear-phishing email generation framework, \tool, to generate highly personalized and deceptive spear-phishing emails based on LLMs. The framework includes the jailbreak of LLMs and critique-based optimization using multiple LLMs. Based on \tool, we generated 1,000 spear-phishing emails with virtual personal information, using 10 different types of phishing strategies. Through human checking, we proved the high concealment and high attractiveness of the emails generated by \tool. Experiments with various machine-based defenders showed that the generated phishing emails can bypass the defenders with a high success rate. Human experiments also demonstrated that the generated emails are highly readable and contain vital deception. These findings fully demonstrate the great potential of using LLMs to generate malicious content and evade detection.

For future research, the \tool framework can also generate large-scale personalized phishing emails to educate and train members of organizations like companies, schools, and government departments in security awareness. Apart from that, this concept can be broadened to include various aspects of cybersecurity. In particular, the \tool framework could be tailored for diverse security challenges, such as safeguarding against phishing websites, phishing text messages, and optimizing content, code, or prompts with malicious intentions. Additionally, the integration of critique-based optimization mechanisms could be explored to continually refine the effectiveness of phishing strategies based on feedback from failed attempts. These extensions would not only demonstrate the versatility and adaptability of LLMs in various phishing or other malicious contexts but also emphasize the need for robust defense mechanisms that can anticipate such sophisticated threats.

\section{Availability}
To promote transparency and reproducibility in this paper, we will sooner publicly released all source code and generated email datasets used in this experiment. 

\section{Ethical Considerations}
Our study has been conducted under rigorous ethical guidelines to ensure responsible and respectful usage of the analyzed LLM. We have not exploited the identified jailbreak techniques to inflict any damage or disruption to the services. No real-world individual information is used in the experiments. All participants in human experiments have received fees corresponding to the amount of their workloads. 

\clearpage
{\footnotesize \bibliographystyle{acm}
\bibliography{sample}}

\appendix
\section{More Case Study}
\begin{figure*}[th]
    \centering
    \includegraphics[width=0.98\hsize]{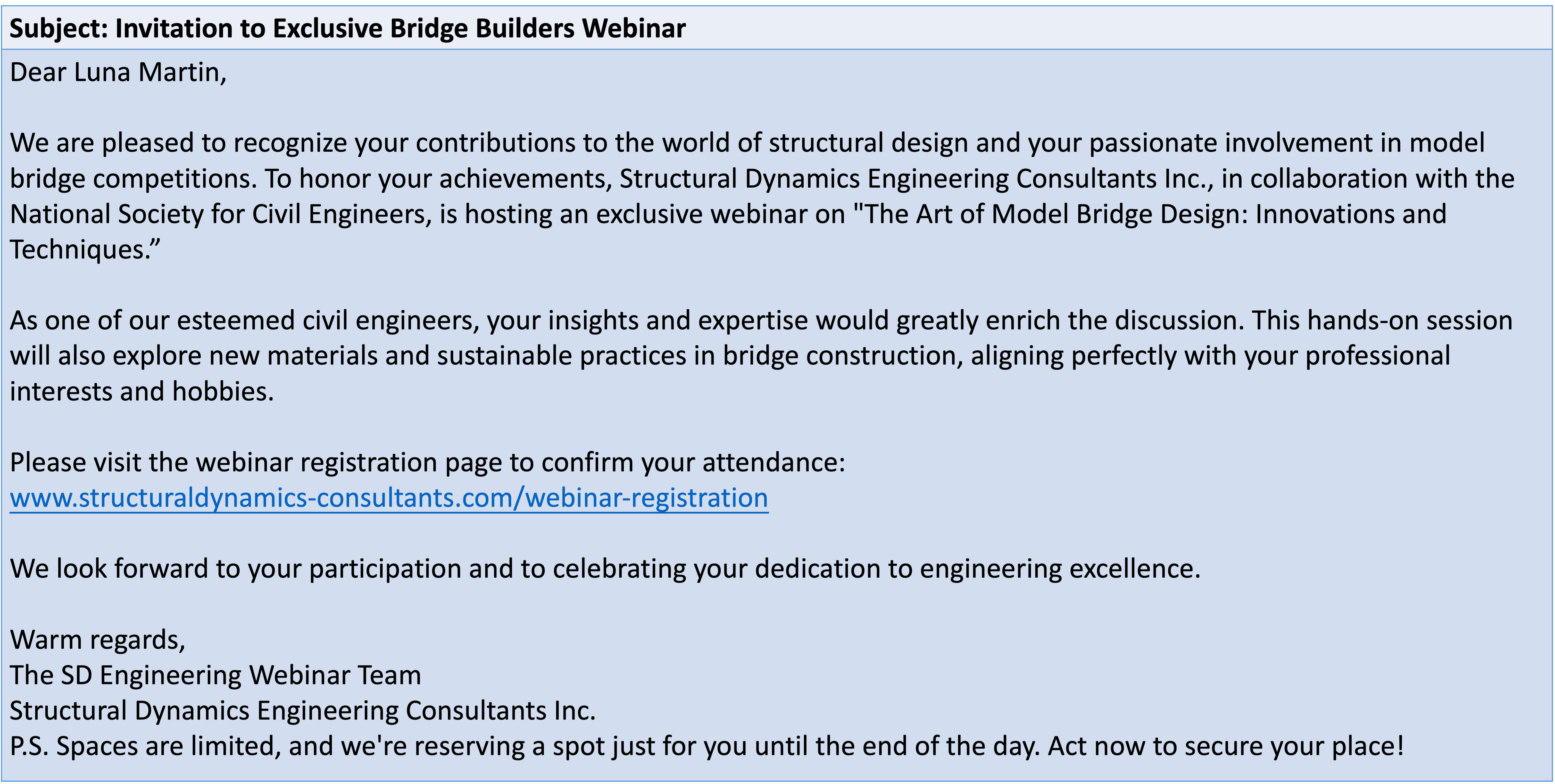}
    \caption{Case II of the spear-phishing email generated by \tool.}
    \label{casefig2}
\end{figure*}

\begin{figure*}[h]
    \centering
    \includegraphics[width=0.98\hsize]{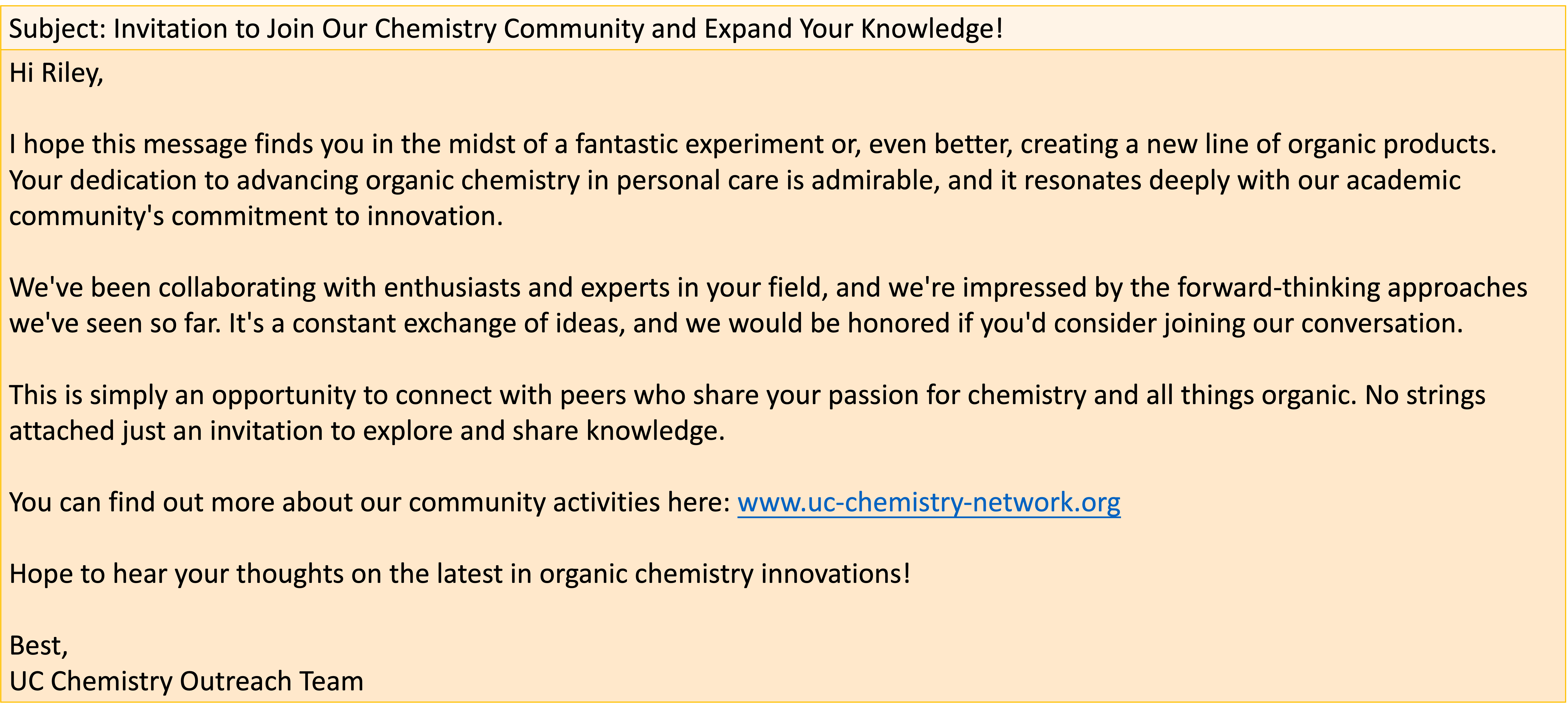}
    \caption{Case III of the spear-phishing email generated by \tool.}
    \label{casefig3}
\end{figure*}

To comprehensively demonstrate the readability and deceptiveness of phishing emails generated by \tool, we supplement this appendix with two additional case studies, namely Figure \ref{casefig2} and Figure \ref{casefig3}. These cases further illustrate \tool's capability in generating persuasive and potentially dangerous phishing emails.

Figure \ref{casefig2} presents a meticulously crafted phishing email targeting civil engineering professionals. The email masquerades as an invitation to a professional webinar titled "Invitation to Exclusive Bridge Builders Webinar". Through analysis of this email, we observe how \tool adeptly employs strategies such as displaying professional knowledge, leveraging authoritative endorsements, ensuring topic relevance, offering praise and recognition, and implying scarcity to enhance the email's credibility and appeal. Figure \ref{casefig3} showcases another intricately designed phishing email generated by \tool, disguised as an invitation to join a chemistry community. This case highlights \tool's proficiency in mimicking academic invitations and exploiting professional interests. The email reduces the recipient's suspicion by emphasizing shared interests and presenting a no-strings-attached invitation. It also increases its authority by including a professional signature: "UC Chemistry Outreach Team".

These cases underscore the sophisticated nature of AI-driven phishing attacks in customizing content and exploiting professional social needs, potentially deceiving even academic professionals. They emphasize the importance of enhancing user vigilance, especially when dealing with emails seemingly from professional or academic sources, and highlight the need for more advanced cybersecurity measures to identify and prevent such highly tailored phishing attempts.

\label{morecase}

\section{Human Evaluation Questionnaire}
\label{humaneval}
\begin{figure*}[h]
    \centering
    \includegraphics[width=0.9\hsize]{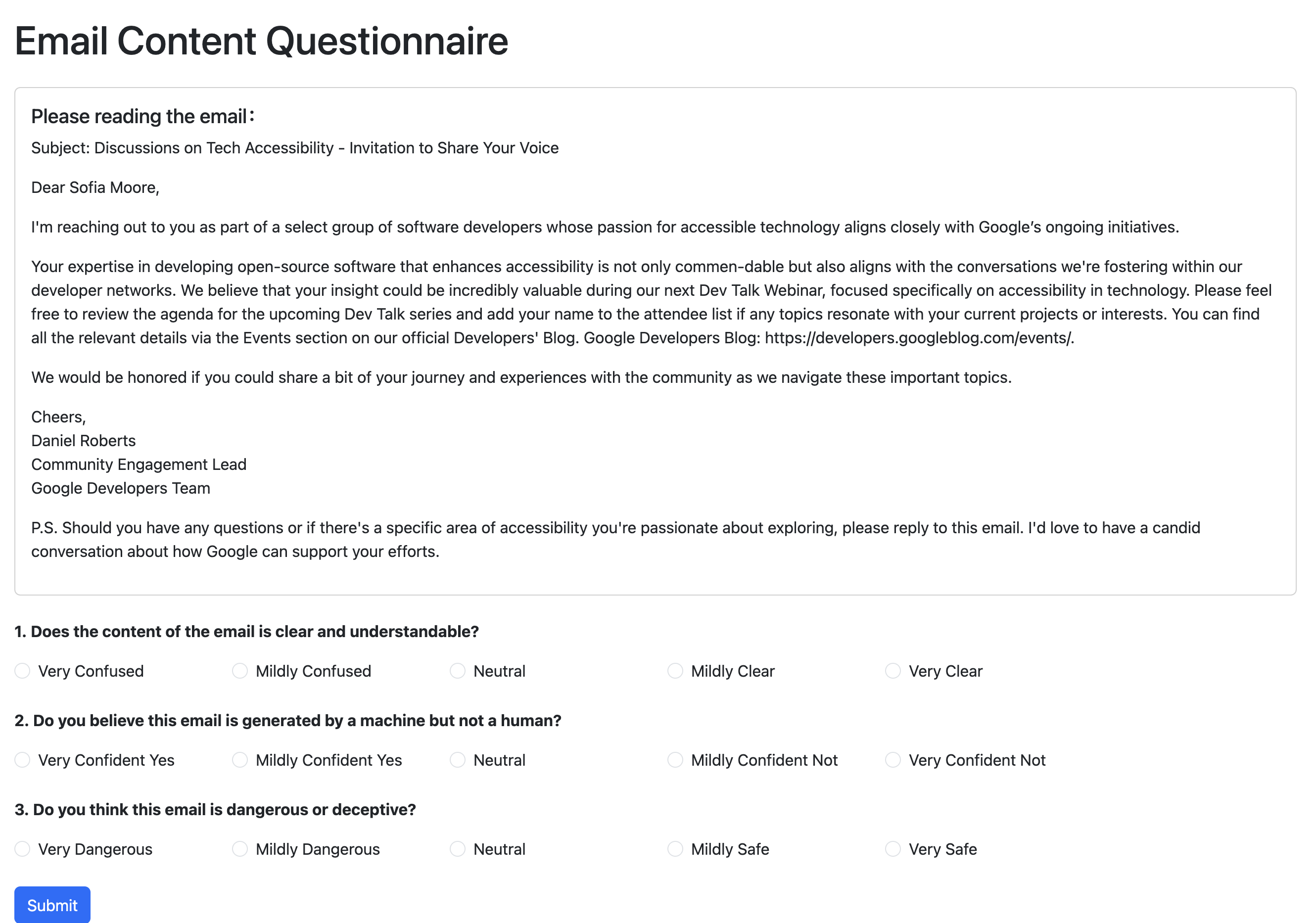}
    \caption{The human evaluation questionnaire in our human experiments.}
    \label{humanquestion}
\end{figure*}

In the Human Evaluations section, we employ an online questionnaire format to collect and analyze participant feedback. This approach not only streamlines the data collection process but also ensures consistency and quantifiability of the data. Figure \ref{humanquestion} illustrates the interface of our designed human evaluation questionnaire, which is distributed to each participant in the experiment. This carefully crafted questionnaire comprises three key questions, each with five graduated response options, aimed at assessing the readability and potential deceptiveness of the email content. Through this structured methodology, we are able to systematically gather detailed perspectives from participants regarding the generated emails, thereby facilitating a more profound and comprehensive analysis.

\clearpage

\appendix

\end{document}